\documentstyle[preprint,aps]{revtex}

\begin{document}
\title{Ground state of excitons and charged excitons in a quantum well}
\author{{\sc C. Riva}\cite{clara} and{\sc \ F. M. Peeters}\cite{francois}}
\address{Departement\ Natuurkunde, Universiteit Antwerpen (UIA), B-2610 Antwerpen.}
\author{{\sc K. Varga}}
\address{Physics Department, Argonne National Laboratories, Argonne,
60439 Illinois.}
\date{\today }
\maketitle

\section{Abstract}

A variational calculation of the ground state of a neutral exciton and of
positively and negatively charged excitons (trions) in a single quantum well
is presented. We study the dependence of the correlation energy and of the
binding energy on the well width and on the hole mass. Our results are
compared with previous theoretical results and with available experimental
data.

\section{Introduction}

Negatively (X$^{-}$) and positively (X$^{+}$) charged excitons, also called 
{\em trions}, have been the object of intense studies in the last years,
both experimentally and theoretically. The stability of charged excitons in
bulk semiconductors was proven theoretically by Lampert\cite{Lampert} in the
\ late fifties, but only recently they have been observed in quantum well
structures: first in CdTe/CdZnTe by Kheng {\it et al.}\cite{Kheng} and
subsequently in GaAs/Al$_x$Ga$_{1-x}$As\cite{Finkelstein,Shields,Manus}.

The calculation we present in this paper, of the ground state energy for the
exciton (X) and the charged exciton, fully includes the Coulomb interaction
among the particles, i.e. no approximated average potential is assumed in
any of the three spatial directions and the correlation among the particles
is fully taken into account.

The Hamiltonian of a negatively charged exciton (X$^{-}$) in a quantum well
is in the effective mass approximation given by 
\begin{equation}
\widehat{H}=T_{1e}+T_{1h}+T_{2e}+V_{C}+V_{1e}+V_{2e}+V_{1h},  \label{HAM}
\end{equation}
where $1e$, $2e$ indicate the electrons and $1h$ the hole; $V_{ie},$ $V_{ih}$
are the quantum well confinement potentials; $T_{i}=\overrightarrow{p}%
_{i}^{2}/2m_{i}$ is the kinetic operator for particle $i$, with $m_{i}$ the
corresponding mass; $V_{C}$ is the sum of the Coulomb electron-electron and
electron-hole interactions, 
\begin{equation}
V_{C}=\frac{e^{2}}{\varepsilon }\left( \frac{1}{|\overrightarrow{r}_{1e}-%
\overrightarrow{r}_{2e}|}-\frac{1}{|\overrightarrow{r}_{2e}-\overrightarrow{r%
}_{1h}|}-\frac{1}{|\overrightarrow{r}_{1e}-\overrightarrow{r}_{1h}|}\right) ,
\end{equation}
with $e$ the elementary charge and $\varepsilon $ the static dielectric
constant. In the present work the heights of the square well confinement
potentials are $V_{ie}=0.57\times (1.155x+0.37x^{2})$ eV for the electrons
and $V_{ih}=0.43\times (1.155x_{0}+0.37x^{2})$ eV for the holes for
the GaAs/Al$_{x}$Ga$_{1-x}$As quantum well system.

The Hamiltonian is then solved using the stochastic variational method\cite
{varga}. The trial function is taken as a linear combination of correlated
Gaussian functions, 
\begin{eqnarray}
&&\phi _{0}(\overrightarrow{r}_{1e},\overrightarrow{r}_{2e},\overrightarrow{r%
}_{1h})=\sum_{n=1}^{K}C_{n0}\Phi _{n0}(\overrightarrow{r}_{1e},%
\overrightarrow{r}_{2e},\overrightarrow{r}_{1h}),  \label{wave-function} \\
&&\Phi _{n0}(\overrightarrow{r}_{1e},\overrightarrow{r}_{2e},\overrightarrow{%
r}_{1h})={}{\cal A}\left\{ \exp \left[ -{\frac{1}{2}}\sum_{%
\renewcommand{\arraystretch}{0.6}{\ 
\begin{array}{c}
\scriptstyle{i,j \in \{1e,2e,1h\}} \\ 
\scriptstyle{k\in \{x,y,z\}}
\end{array}
}}A_{nijk0}r_{ik}r_{jk}\right] \right\} ,
\end{eqnarray}
where $r_{ik}$ gives the positions of the $i-th$ particle in the direction $%
k $; ${\cal A}$ is the antisymmetrization operator and $\{C_{n0},A_{nijk0}\}$
are the variational parameters. The dimension of the basis, $K$, is
increased until the energy is sufficiently accurate.

\section{The results}

\noindent The correlation energy of a charged exciton is defined as 
\begin{eqnarray}
E_{C}(X^{-}) &=&E_{T}(X^{-})-2E_{e}-E_{h}, \\
E_{C}(X^{+}) &=&E_{T}(X^{+})-2E_{h}-E_{e},
\end{eqnarray}
with $E_{T}(X^{\pm })$ the energy level of the charged exciton and, $E_{e}$
and $E_{h}$ the energy levels of the free electron and hole, respectively,
in the quantum well. We discuss here the results obtained for a GaAs/Al$_{x}$%
Ga$_{1-x}$As quantum well with $x=0.3$. The values of the GaAs masses used
are $m_{e}=0.0667m_{0},$ $m_{hh}=0.34m_{0},$ which results into $%
2R_{y}=\hbar /m_{e}a_{B}=11.58$ meV and $a_{B}=\hbar ^{2}\varepsilon
/m_{e}e^{2}=99.7$ \AA .

Our numerical results for the correlation energy are shown in Fig. \ref
{Comp-stebe} and are compared with the results of Ref. \onlinecite{Stebe97}.
For the X we find that the magnitude of the correlation energy is larger
than the one obtained in Ref. \cite{Stebe97} while for the X$^{-}$ our
approach gives a $5\%$ smaller magnitude for the correlation energy$.$ The
reasons for the difference are: 1) in Ref. \onlinecite{Stebe97} the Coulomb
potential along $\ $the $z$-direction was approximated by an analytical form, see Appendix in Ref. \cite{Stebe97}. This
approximation leads, as the authors already noted, to an error in the X
correlation energy of approximately 5\%; 2) for the X$^{-}$ energy the authors
of Ref.\cite{Stebe97} did not report an estimate of the error introduced
by the approximations made. However we find a decrease of the absolute
value of the correlation energy by about 5\%. We think that this result is
not in conflict with the one obtained for the X. In fact if in Ref. \cite
{Stebe97} for the X case the intensity of the attractive interaction between
the electron and hole (e-h) was underestimated, for the X$^{-}$ case also
the intensity of the repulsive interaction between the electrons (e-e)
is underestimated which leads to a less negative E$_{C}$.

We also report the correlation energy for the X$^{+}$ in Fig. \ref
{Comp-stebe}. Note that the correlation energy of the X$^{+}$ is practically
equal to the one of X$^{-}$. This is in perfect agreement with recent
experimental data \cite{Finkelstein} where the binding energy of the X$^{+}$
is found to be equal to the one of \ the X$^{-}$.

Next we take into account, for the narrow well regime, the difference in
mass of the particles in the well (GaAs) and in the barrier (Al$_{x}$Ga$%
_{1-x}$As) material. The values for the GaAs-masses, i.e. the masses for the
electron and the hole, are taken equal to the one used in the previous
calculation. The values for the masses in Al$_{x}$Ga$_{1-x}$As are $%
m_{eb}^{\ast }=0.067+0.083x,$ $m_{hhb}^{\ast }=0.34+0.42x,$ \noindent where $%
x$ indicates the percentage of Al present in the alloy. If we assume, as a
first approximation, that the electron and the hole have part of their wave
function in the quantum well and the rest in the barrier we may take the
total effective mass of the electron and the hole as given by 
\begin{equation}
\frac{1}{m_{i}}=\frac{P_{iw}}{m_{iw}}+\frac{P_{ib}}{m_{ib}},
\end{equation}
where $m_{iw},m_{ib}$ are the masses of the $i$-th particle in the barrier
and in the well, and $P_{iw}$, $P_{ib}$ are the probabilities of finding the 
$i$-th particle in the well or in the barrier, respectively. The results of
this calculation are shown in Fig. \ref{asymm}. We observe that the effect
of the mass mismatch is important only in the narrow quantum well regime,
i.e. L$<$ $40$ \AA , where it leads to a substantial increase of the energy.

The dependence of the total energy on the hole mass for a 200 \AA\ \ wide
quantum well is shown in Fig. \ref{boh}. The energy of the negatively
charged exciton becomes equal to the D$^{-}$ energy\cite{clara-d} for $%
m_{h}/m_{e}>$ 16. Note that the X$^{+}$ energy is for large values of the
hole electron mass ratio parallel to the one of the X$^{-}$. For a large
hole mass its contribution to the total energy in terms of confinement
energy is negligible, and the difference between the total energy of the
positively and the negatively charged exciton is just due to the confinement
energy of one electron, which does not dependent on the hole mass.

Experimental data were reported for the binding energy of the X$^{-}$ in
zero magnetic field for a 200 \AA \cite{Finkelstein}, a 220 \AA \cite
{Shields95}\ and a 300 \AA \cite{Shields95}\ quantum well. The binding
energy is defined as $E_{B}=E_{T}(X)+E_{e}-E_{T}(X^{-}).$ The
experimental results are shown in Fig. \ref{binding} together with our
theorethical calculation, where the shaded band indicates the estimate
accuracy of our variational procedure. Notice that the experimental results
give a larger binding energy as compared to the theoretical estimate and
this discrepancy increases with decreasing well width. This may be a
consequence of the localization\cite{oliver} of the trion due to well width
fluctuations which become more important with decreasing L.

Last we study the wave function of the X$^{-}$. In Fig. \ref
{geometry}(a) we show the contour plot of $|\phi _{0}(%
\overrightarrow{r}_{1e},\overrightarrow{r}_{2e},\overrightarrow{r}_{1h})|^{2}
$ for a X$^{-}$ in a quantum well of width 100 \AA. We fix the hole in $\overrightarrow{r}_{h}=(0,0,0)$ and one of the
two electrons in$\overrightarrow{r}_{e}=(-0.25a_{B},0,0)$ and calculate the
probability to find the other electron in the $\widehat{xy}$-plane. Notice
that the second electron sits far from the hole and the fixed electron and
behaves like an electron weakly bound to a polarized exciton.
If now we fix the hole in $\overrightarrow{r}_h=(2a_B,0,0)$ and the
electron in $\overrightarrow{r}_e=(0,0,0)$, see
Fig. \ref{geometry}(b), we observe that the second electron is
completely localized around the hole and the configuration that we
obtain is the one of an exciton plus an extra electron.

\section{Conclusion}

In this paper a new calculation for the exciton and the charged exciton
energy in a quantum well was presented which is based on the stochastic
variational method. To our knowledge, this is the first time, that a
calculation fully includes the effect of the Coulomb interaction and the
confinement due to the quantum well. The results obtained do not show a big
qualitative difference from the one already present in the literature,
however a sensible quantitative difference is observed. This difference
leads to an improvement of the agreement with available experimental data
for the binding energy.

\section{\protect\bigskip Acknowledgment}

Part of this work is supported by the Flemish Science Foundation (FWO-Vl)
and the `Interuniversity Poles of Attraction Program - Belgian State, Prime
Minister's Office - Federal Office for Scientific, Technical and Cultural
Affairs'. F.M.P. is a Research Director with FWO-Vl. Discussions with M.
Hayne are gratefully acknowledged.

\begin{figure}[tbp]
\caption{The correlation energy vs. the quantum well width for the heavy
hole exciton and charged exciton. }
\label{Comp-stebe}
\end{figure}
\begin{figure}[tbp]
\caption{The correlation energy of the negative charged exciton vs. the
quantum well width. For the case of constant masses, in the hole and in the
barrier, and for the the case of different masses.}
\label{asymm}
\end{figure}
\begin{figure}[tbp]
\caption{The total energy of the negative charged exciton and the positive
charged exciton vs. $m_{h}/m_{e}$ for a quantum well of width 200
\AA\. The total energy of a D$^{-}$ in the same quantum well is shown
(dotted line) for comparison.}
\label{boh}
\end{figure}
\begin{figure}[tbp]
\caption{The theoretical and experimental binding energies of the negatively
charged exciton vs. the well width.}
\label{binding}
\end{figure}
\begin{figure}[tbp]
\caption{Contour maps for the conditional probability of the electron  in X$%
^{-}$. The symbols indicate the
position of the fixed particles.}
\label{geometry}
\end{figure}

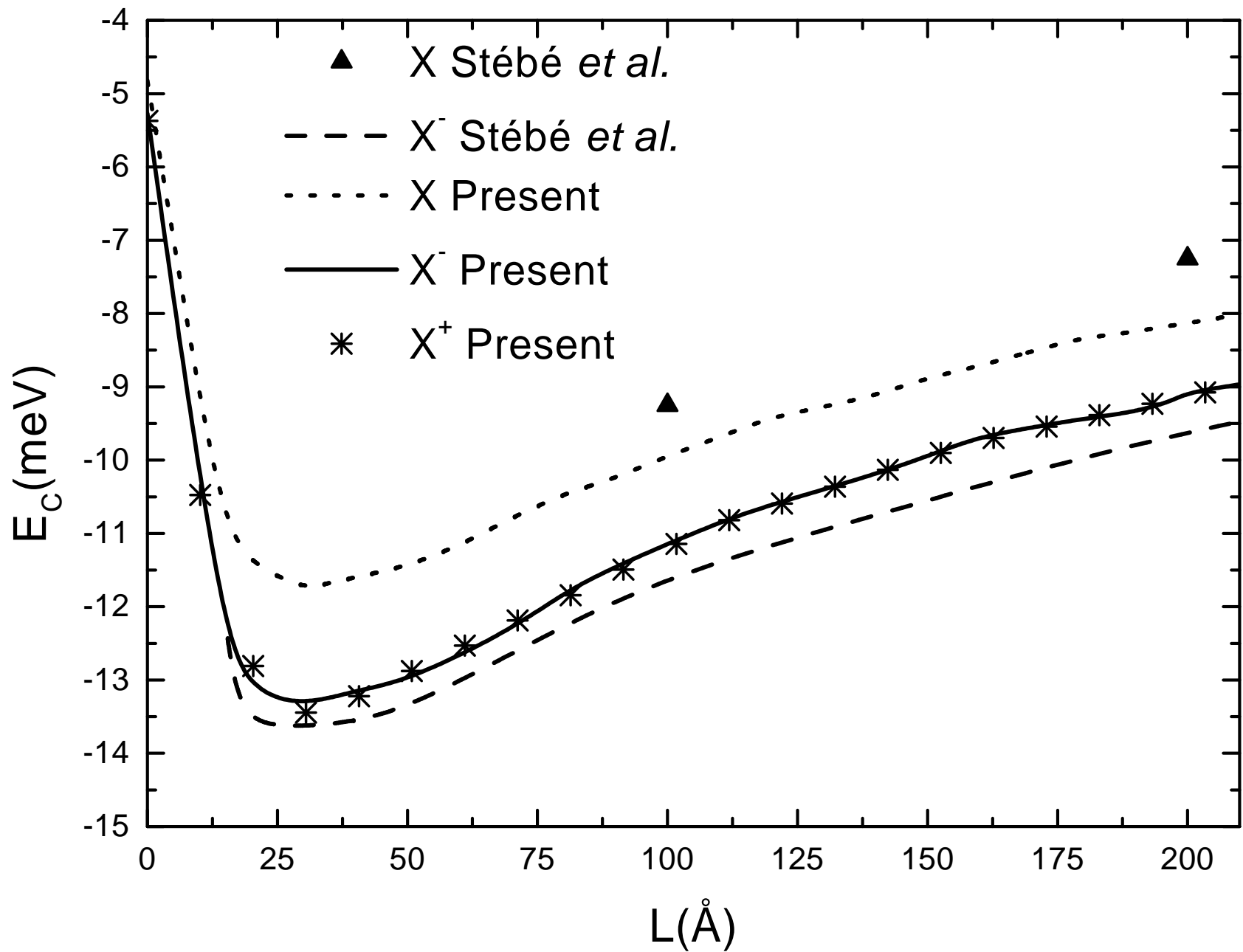
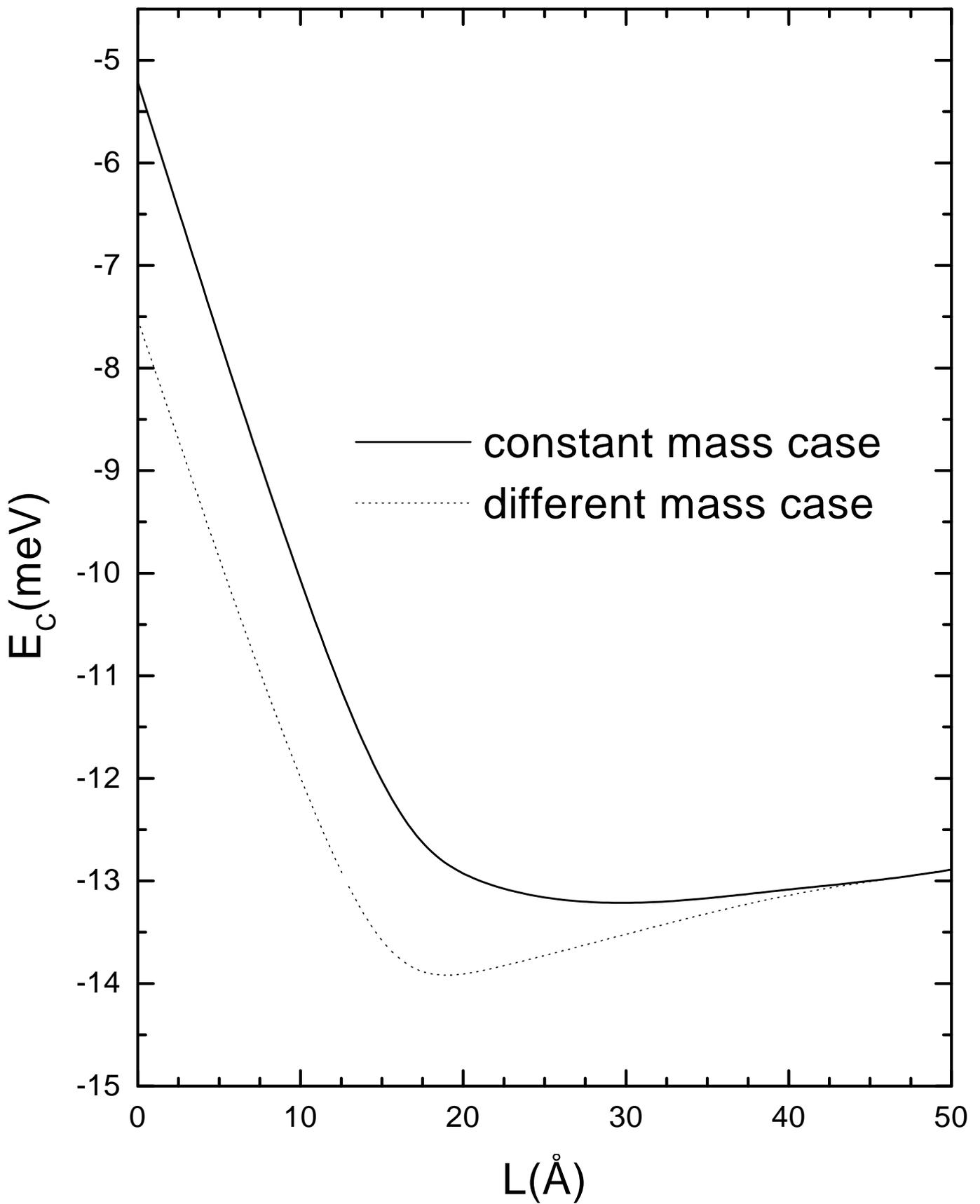
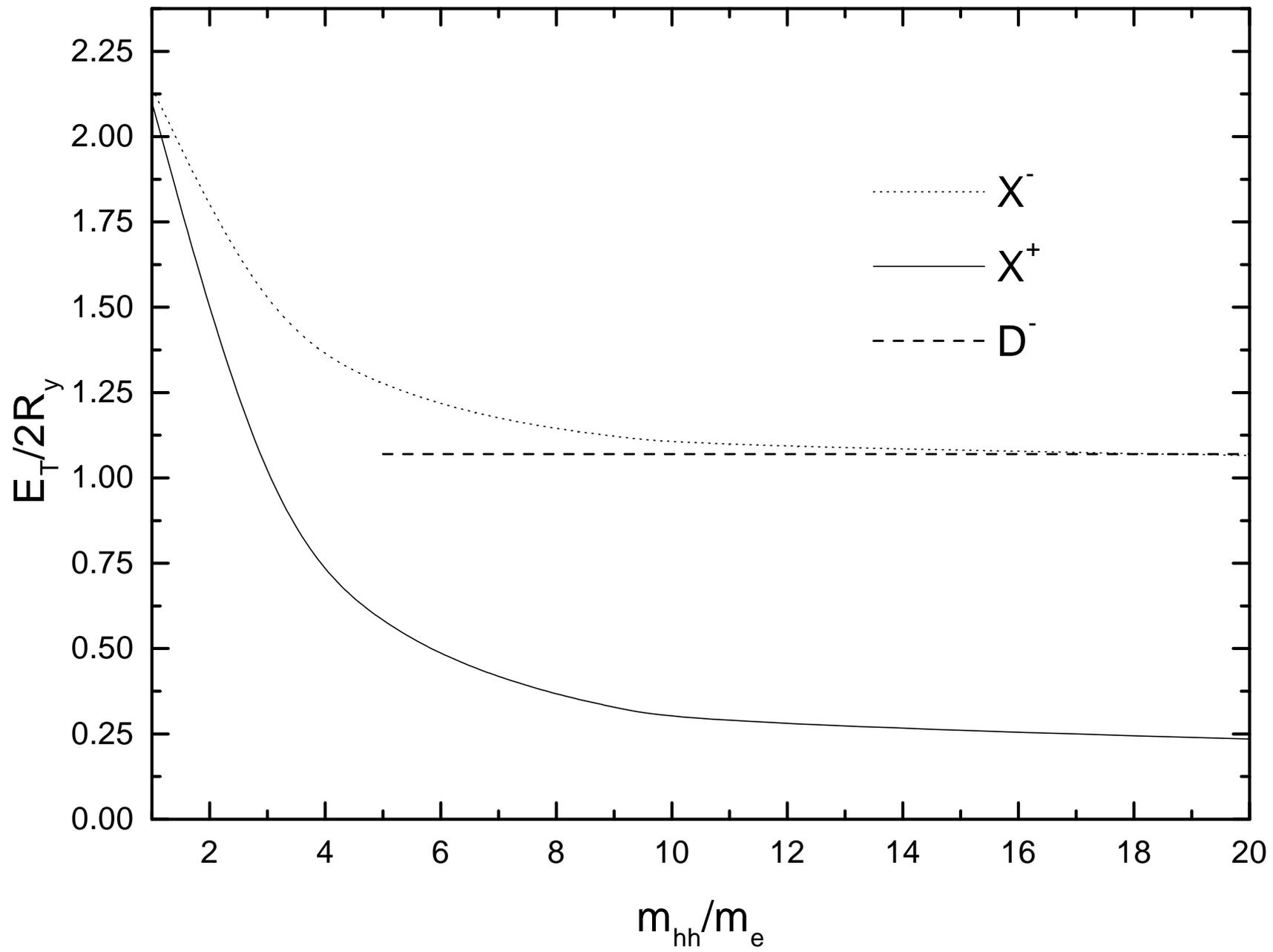
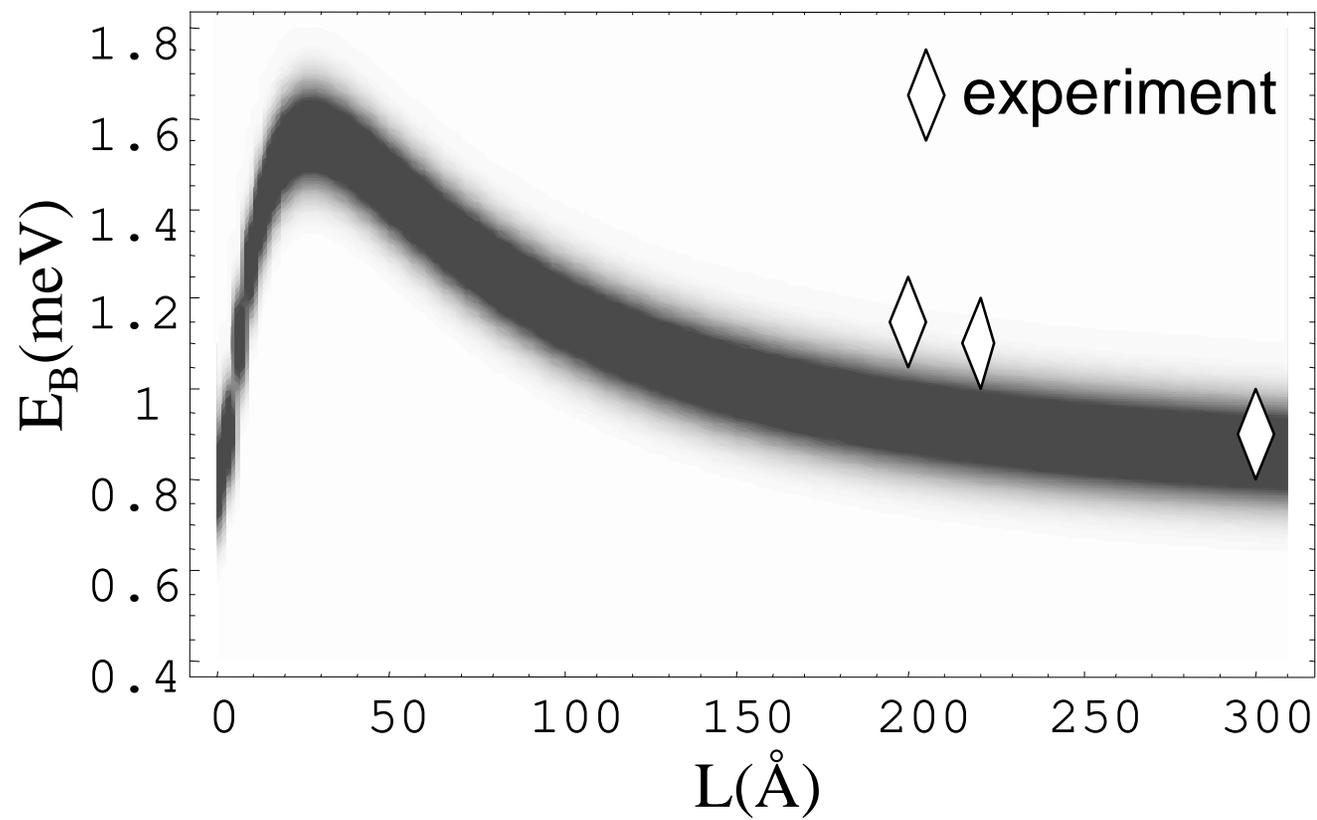
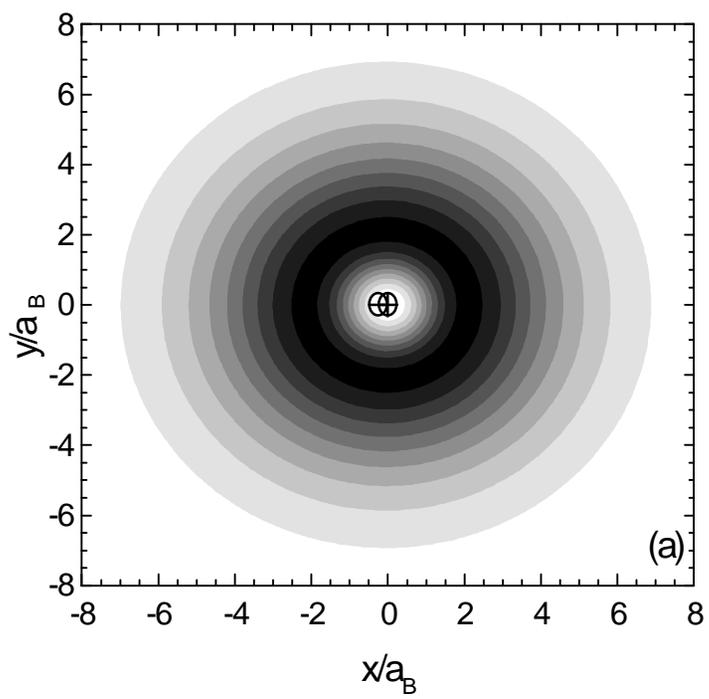
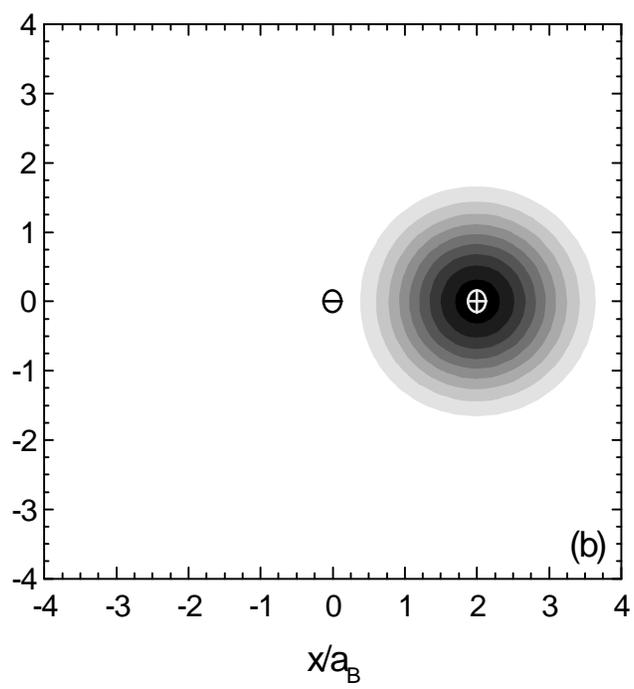
\end{document}